\documentclass[12]{article}
\usepackage{amssymb}

\newtheorem{theorem}{Theorem}

\newtheorem{corollary}{Corollary}

\title{Bell's Inequalities and the Accardi--Gustafson Inequality\footnote{To be
presented at the
International Conference on Foundations of Probability and Physics---2,
V\"axjo, Sweden, June 2--7, 2002.}}
\author{Karl Gustafson\\
Department of Mathematics, University of Colorado,\\
Boulder, CO 80309--0395 USA}

\begin{document}
\maketitle
\begin{abstract}
Many issues combine for consideration when speaking of Bell's
Inequalities:  nonlocality, realism, hidden variables, incompatible
measures, wave function collapse, other.  Each of these issues then may
be viewed from several viewpoints:  historical, theoretical, physical,
experimental, statistical, communicational, cryptographical, and
mathematical.  From the mathematical viewpoint, much
of the Bell theory is ``just geometry''.

\begin{theorem}
Let $x,y,z$ be any 3 nonzero vectors in a real or complex Hilbert space
of any dimension.  We take $\Vert x\Vert = \Vert y\Vert = \Vert z\Vert =
1$ for convenience.  From $\langle x,y\rangle = a_1 +ib_1$, $\langle
y,z\rangle = a_2 +ib_2$, $\langle x,z\rangle = a_3 +ib_3$, define angles
$\phi_{xy}, \phi_{yz}, \phi_{xz}$ in $[0,\pi]$ by $\cos \phi_{xy} = a_1$,
$\cos_{yz} = a_2$, $\cos \phi_{xz} = a_3$.  Then there holds the general
triangle inequality
$$\phi_{xz} \leqq \phi_{xy} + \phi_{yz} $$
\end{theorem}

\begin{corollary}
Much of the Bell (1965), Wigner (1970), Accardi (1982), Gudder--Zanghi (1984),
Herbert--Peres (1993), Williams--Clearwater (1998), Khrennikov (2000), other,
considerations are contained in the above Theorem.
The Accardi--Gustafson
inequality is a necessary condition for a quantum mechanical probability
model to apply to those situations.
From the mathematical viewpoint of this paper, one cannot argue
``nonlocality'' on the basis of violation of Bell's Inequality.
\end{corollary}
\end{abstract}

\section{Introduction and Outline}
In this report I want to summarize and bring up to date my results/impressions
of the last five years about what I loosely call the Bell theory.  Although I
have been interested in the foundations of quantum mechanics for a long time I
specifically came to the Bell theory from my observing about five years ago a
coinciding of certain inequalities in quantum probability with those from my
operator trigonometry.  My operator trigonometry started in abstract operator
theory in 1967, is inherently noncommutative, and includes somewhat incidentally
the triangle inequality mentioned in the abstract above.  I will emphasize the
mathematical viewpoint and often defer to the cited literature 
\cite{einstein1}--\cite{gustto} so as to
not repeat myself/others in this limited space.

In Section 2 I briefly review the Bell theory and some of Bell's inequalities.
In Section 3 I detail the above mentioned coincidence of what I shall call here
the Accardi--Gustafson inequality, which is essentially equivalent to the
triangle inequality but in my opinion is even more fundamental as I shall try to
make clear.  In Section 4 I will introduce a new notion of ``inequality
equalities'' which I believe will be useful in revealing flaws in paradoxes.
Section 5 contains concluding remarks which may be of some interest in
themselves.

\section{Bell's Inequalities}
Here is a thumbnail sketch of how Bell's inequalities fit into the foundations
of quantum mechanics.  As with the term Bell theory, I also use the term Bell's
inequalities in a wide sense, meaning many such inequalities.

The 1935 paper \cite{einstein1} of Albert Einstein, Boris Podolsky and
Nathan Rosen was a {\it gedankenexperiment} which purported to
demonstrate that quantum mechanics cannot provide a complete description
of reality.  According to the extensive account \cite{jammera} of Jammer,
although much of the actual EPR paper \cite{einstein1} was written by
Podolsky, the origins of this paper go back to 1930 when Niels Bohr
`defeated' Einstein's earlier {\it gedankenexperiment} presented at the
Sixth Solvay Congress in Brussels in 1930, an important episode in in the
famous ongoing debate between the two which had begun already ten years
earlier in 1920.  To better present his view, Einstein then, along with
Richard Tolman and Podolsky, wrote a paper \cite{einstein2} with another
{\it gedankenexperiment} which argued that if one accepted quantum
mechanic's uncertainty principle, then one could not even predict the
past, let alone the future.  Also Einstein sharpened his arguments by
shifting attention away from direct attacks on the uncertainty principle
itself but instead with more focus on logical paradoxes which would
follow from it \cite{einstein3}.  Moreover Einstein modified his photon
box {\it gedankenexperiments} with their paradoxical consequences to the
more clear-cut two particle {\it gedankenexperiment} which appears in the
1935 EPR paper \cite{einstein1}.  Erwin Schr\"odinger immediately agreed
with the EPR argument, reformulated it, and came up with his own {\it
gedankenexperiment} now known as his half-dead half-alive cat
\cite{schrod}.

The conclusion of the groundbreaking paper\cite{einstein1} was:  
``While we have thus
shown that the wave function does not provide a complete description of
the physical reality, we left open the question of whether or not such a
description exists.  We believe, however, that such a theory is possible.''
Thus the emphasis in [1] was on inadequacies of a theory in which all
information is in the wave function.  In 1951 David Bohm \cite{bohm1}
responded by reformulating the EPR argument to one expressed more simply
in terms of spin functions, and presented an argument that ``no theory of
mechanically determined hidden variables can lead to all of the results
of the quantum theory.''  Nonetheless Bohm then introduced his version of
such a hidden variable theory.  This was like earlier semiclassical
hydrodynamical or pilot wave quantum models, except for two new features.
First, the existence of a quantum mechanical potential, shall we say among
all of the particles in a considered ensemble $\psi$, was assumed.
Second, each particle trajectory will be well-posed if you know its
initial condition.  But because the initial position could not be
experimentally measured, it is a hidden variable.

In 1964 Bell \cite{bell64} presented his famous inequality 
$$|P(a,b)-P(a,c)| \leqq 1 + P(b,c) \eqno(2.1)$$
and exhibited
certain quantum spin measurement configurations whose quantum expectation
values could not satisfy his inequality.  Bell's analysis assumes that
physical systems, e.g. two measuring apparatuses, can be regarded as
physically totally separated, in the sense of being free of any effects
one from the other.  Thus his inequality could provide a `test' which
could be failed by measurements performed on correlated quantum systems.
In particular it was argued in [7] that local realistic hidden variable
theories could not hold.  However, the exact nature of hidden variables
as viewed by Bell is unclear from [7].
As is well known the 1982 physical experiments of Aspect {\it et al.}
\cite{aspect} 
demonstrated that beyond any reasonable doubt the Bell inequalities are
violated by certain quantum systems, and papers continue to appear with
further demonstrated violations.  

In a 1970 paper \cite{wigner}, Wigner simplified and clarified in 
several ways the
argument of Bell.  Wigner assumed that all possible measurements are
predetermined, even if they involve incompatible observables, and
moreover any measurement on one of two apparatuses does not change the
preset outcomes of measurements on the other apparatus.  Thus the
meanings of locality and realism are made more clear and both assumptions
are present in the model setup.  It is helpful to imagine, for example,
that the `hidden variable' is just the directional orientation of each of
the two apparatuses, each of which can be thought of as just a
three-dimensional possibly skew coordinate system, for example.  Then two
spin 1/2 particles are sent to the apparatuses, each to one, both coming
from a common atomic source, with perfect anticorrelation and singlet
properties.  Nine measurements are then needed to simultaneously measure
the direction vectors $\omega_1 ,\omega_2 ,\omega_3$ of the two spins.
Each spin has two possible values $1/2 \equiv +$, $-1/2 \equiv -$, so
each measurement can permit four relative results:  $++$, $--$, $+-$, $-+$.
Therefore there are $4^9$ possible outcomes.  Wigner then assumes that
the spins are not affected by the orientation of the particular measuring
apparatus.  This reduces the outcomes to $2^6$ possibilities.  For
example, if the hidden variables are in the possibility domain $(+, -,
-,; - + -)$, then the measurement of the spin component of the first
particle in the $\omega_1$ direction will yield value spin $= +$, no
matter what direction the spin of the second particle is measured.

In the above setup, let $\theta_{12} ,\theta_{23} ,\theta_{31}$ be the
angles between the three directions $\omega_1 ,\omega_2 ,\omega_3$.  Then
the probability that the spin component of particle 1 in the $\omega_i$
direction and the spin component of particle 2 in the $\omega_k$
direction both measure + or both measure $-$ is $\frac{1}{2} \sin^2
\left( \frac{\theta_{ik}}{2} \right)$.  Otherwise the probability of
measurements $+ -$ or $- +$ is $\frac{1}{2} \cos^2 \left(
\frac{\theta_{ik}}{2} \right)$.  I mention that even though quantum mechanics
has been statistically interpreted ever since Born in 1927, the fact that such 
quantum transition probabilities are
given trigonometrically as $|\langle \psi (\omega_i ), \psi (\omega_k
)\rangle |^2 = \cos^2 (\theta_{ik} /2)$ in terms of the angle
$\theta_{ik}$ between directions $\omega_i$ and $\omega_k$ is a special
property of spin systems and for example the Eulerian angle
representation for $SU(2)$, see, e.g. (\cite{miller}, p. 225).  Wigner's version of
Bell's theory then becomes the inequality
$$\frac{1}{2} \sin^2 \frac{1}{2} \theta_{23} + \frac{1}{2} \sin^2 \theta_{12}
\geqq \frac{1}{2} \sin^2 \frac{1}{2} \theta_{31} \eqno(2.2)$$

There are many other versions of Bell's inequalities.  See the literature
\cite{jammer}--\cite{bohm2}
for example.  I will only consider one more of them in this paper.
Consider the nice treatment of the important CHSH \cite{clauser} inequality 
in Bohm (\cite{bohm2},
pp. 347--354).  A very large number of particles in the spin singlet state
are considered.  Let ${\mathbf a}, {\mathbf b}, {\mathbf c}, {\mathbf 
d}$ be four
arbitary chosen unit vector directions in the plane orthogonal to the two
beams produced by the source.  Let $v_i ({\mathbf a})$ and $v_i ({\mathbf
d})$ be the ``hidden'' predetermined values $\pm 1$ of the spin
components along {\bf a} and {\bf d}, respectively, of particle 1 of the
$i$th pair, similarly $w_i ({\mathbf b})$ and $w_j ({\mathbf c})$ for
particle 2 values along directions {\bf b} and {\bf c}.  Then the average
correlation value for particle 1 spins measured along {\bf a} and
particle 2 spins measured along {\bf b} is
$$E({\mathbf a}, {\mathbf b}) = \frac{1}{N} \sum_{i=1}^N v_i ({\mathbf 
a}) w_i ({\mathbf b}) \eqno(2.3)$$
In the same way one considers the average correlation values $E({\mathbf
a},{\mathbf c})$, $E({\mathbf d},{\mathbf b})$, $E({\mathbf d},{\mathbf
c})$ and adding up all pairs as $i$ runs from 1 to $N$ one arrives at the
CHSH inequality
$$|E({\mathbf a},{\mathbf b}) + E({\mathbf a},{\mathbf c}) + E({\mathbf
d},{\mathbf b}) - E({\mathbf d},{\mathbf c})| \leqq 2. \eqno(2.4)$$
Demanding this estimate hold as well for quantum mechanical expectations
$E({\mathbf a},{\mathbf b}) = -{\mathbf a}\cdot{\mathbf b}$, one has
(\cite{bohm2}, p. 349)
$$\begin{array}{ll}
|{\mathbf a}\cdot{\mathbf b} + {\mathbf a}\cdot{\mathbf c} + {\mathbf
d}\cdot{\mathbf b} - {\mathbf d}\cdot{\mathbf c}| &= |{\mathbf a}\cdot
({\mathbf b}+{\mathbf c}) + {\mathbf d}\cdot ({\mathbf b}-{\mathbf c}|\\
[1em]
&\leqq |{\mathbf a}| |{\mathbf b}+{\mathbf c}| + |{\mathbf d}||{\mathbf
b}-{\mathbf c}|\\
[1em]
&= \sqrt{2+2\cos \phi} + \sqrt{2-2\cos \phi}\end{array} \eqno(2.5)$$
where $\phi$ is the angle $\theta_{bc}$ (a notation I will use later)
between {\bf b} and {\bf c}.  Then one observes that the last expression
is maximized to value $2\sqrt 2$ when $\theta_{bc} = \pi /2$, and ``any
configuration sufficiently `near' to'' the directions providing this
maximal violation of Bell's inequality will also violate it.

\section{The Accardi--Gustafson Inequality}
Here is a brief account of a partial coincidence of two distinct theories, one
operator--theoretic, the other quantum--probabilistic, from which emerged what I
will call here the Accardi--Gustafson inequality.  About five years ago I noticed
this coincidence.  More details on the related work by Accardi et al. in quantum
probability may be found in \cite{accardi1}--\cite{accardi4} and later papers.  
For the operator
trigonometry see the early papers \cite{gust68a}--\cite{gust77} and the recent 
books and surveys \cite{gust97}--\cite{gust01a}.  
More details on the operator trigonometry, as it applies to quantum
probabilities may be found in the recent papers \cite{gust99b}--\cite{gust01b}.

What I noticed was that that Accardi and Fedullo's
(\cite{accardi2}, Proposition 3, Eq. (19)), namely
$$\cos^2 \alpha + \cos^2 \beta + \cos^2 \gamma - 1 \stackrel{<}{=} 2 \cos
\alpha \cos \beta \cos \gamma \eqno(3.1)$$
a necessary and sufficient condition for the angles $\alpha ,\beta
,\gamma$ of a quantum spin model in a 2-dimensional complex Hilbert
space, is precisely the same as (see, e.g., Gustafson and Rao
\cite{gusra}, Lemma 3.3-1, equation (3.3-3)) the operator trigonometry
relation
$$1 - a_1^2 - a_2^2 - a_3^2 + 2a_1 a_2 a_3 \stackrel{>}{=} 0 \eqno(3.2)$$
for the real cosines $a_1 ,a_2 ,a_3$ of the angles between arbitrary unit
vectors in any Hilbert space.  The angles of inequality (3.1) are
related to transition probability matrices $P(A \mid B), P(B \mid C), P(C
\mid A)$ for three observables $A, B, C$ which may take two values.  The
angles of (3.2) are related to a triangle inequality for general operator
angles within the general operator trigonometry.  I have stated that triangle
inequality in the Abstract as Theorem 1.  As indicated in Corollary 1, in my
opinion many ``Bell inequalities'' and related inequalities are best seen in the
light of that general triangle inequality.

Although the triangle inequality of Theorem 1 gives the geometrical manings, I
have come to the opinion that the Accardi--Gustafson inequality (3.1), (3.2) is
more fundamental and certainly more useful.  Because I have discussed elsewhere
\cite{gust99b}--\cite{gust01b}
the application of the operator trigonometry to the quantum spin
probabilistic settings, let me not repeat myself here.  Instead I would like to
present some new or previously unmentioned observations.

I discovered the general vector angle triangle inequality
$$\phi_{xz} \leqq \phi_{xy} + \phi_{yz} \eqno(3.3)$$
in 1966 as I was creating what came to be the operator trigonometry.  However my
chief interest was in the operator version 
$$\sin \phi (B) \leqq \cos \phi (A)\eqno(3.4$$
which becomes a sharp sufficient
condition for the product $BA$ of two bounded noncommuting positive selfadjoint
operators $A$ and $B$ to be accretive (i.e., $\mbox{Re} \,BA$ is positive).
This question came out of abstract Hille--Yosida operator semigroup theory
\cite{gust68a}--\cite{gust68c}.  The definitions of the entities in (3.4) are
$$\sin \phi (B) = \inf_{\epsilon > 0} \Vert \epsilon B-I\Vert , \quad \cos \phi
(A) = \inf_{x\not= 0} \frac{\langle Ax,x\rangle}{\Vert Ax\Vert\Vert x\Vert}
\eqno(3.5)$$
A key early result was the min-max theorem \cite{gust68d} for any positive 
selfadjoint operator $B$
$$\sin^2 \phi (B) + \cos^2 \phi (B) = 1 \eqno(3.6)$$
In my opinion one has no operator trigonometry without this result.  I mention
that (3.6) generally fails when you depart Hilbert space.

In connection with this positive operator product question my student D. Rao and
I found that Krein \cite{krein} also had (slightly later) written down a version
of (3.3) for another
purpose.  Many years later (1995) someone (H. Schneider) pointed out to me that
Wielandt \cite{wielandt} had also written down (slightly later) another verson.  
See the 
historical account \cite{gust96}.  Thus the operator--trigonometric origins of 
the Accardi--Gustafson inequalities predate those of the quantum probability.

The proof of (3.3) can be either elegant or constructive.  The elegant proof is
the following.  None of me, Krein, Wielandt wrote it down.  Take the 3 unit
vectors $x$, $y$ and $z$ and embed them in real 3-space.  Then great circle
spherical distance is a (Riemannian) metric equivalent to the angles.  However,
I also wanted a constructive proof and Rao and I worked out a version in his
dissertation \cite{rao}, \cite{gust77}.  We used the Gram matrix
$$G = \left[ \begin{array}{lll}
\langle x,x\rangle &\langle x,y\rangle &\langle x,z\rangle\\
[1em]
\langle y,x\rangle &\langle y,y\rangle &\langle y,z\rangle\\
[1em]
\langle z,x\rangle &\langle z,y\rangle &\langle z,z\rangle \end{array}
\right] \eqno(3.7)$$
A Gram matrix is positive semidefinite in any number of dimensions, and
definite iff the given vectors are linearly independent.  From it one has in the
present situation the determinant
$$G = \left| \begin{array}{lll}
1 &a_1 &a_3\\
a_1 &1 &a_2\\
a_3 &a_2 &1\end{array} \right| = 1 + 2a_1 a_2 a_3 - (a_1^2 + a_2^2 + a_3^2)
\geqq 0 \eqno(3.8)$$
which is the Accardi--Gustafson inequality (3.1), (3.2).  As I showed in
\cite{gust99b}--\cite{gust01b},
from (3.8) one may constructively quickly prove the triangle
inequality (3.3).  It is a little harder to go the other way.  That is why I
regard the Accardi--Gustafson inequality as more fundamental.

Gudder--Zangli \cite{gudder} also give a relative simple proof of the 
Accardi--Fedullo\cite{accardi2}
inequality (3.1) without using a Grammian and without noting the triangle
inequality (3.3).  Others, e.g. Accardi--Fedullo [26], Wigner \cite{wigner}, 
Williams and Clearwater \cite{willi},
Herbert and Peres \cite{herbert}, \cite{peres}, Khrennikov \cite{khren} 
allude to
resemblances or analogies to triangle inequalities in their work but the value
of (3.3) is that it {\it is} the underlying triangle inequality.  On the other
hand I believe the Accardi--Gustafson inequality (3.1), (3.2), (3.8) is more
fundamental not only in the sense mentioned above but also because in principle
we may use the Grammian for any number of vectors $x,y,z,w,\dots$.  That could
yield geometric--trigonometric relationships in higher dimensions that would be
valuable for quantum probability in the future.

\section{Inequality Equalities}
In \cite{gust00}--\cite{gust01b} I introduced the notion of violation 
boundaries for Bell-like
inequalities and I worked out the theory of such for the Bell inequality (2.1).
The idea was to try to embed such a given inequality into the operator
trigonometry so as to determine exactly where the inequality is violated.
Because the Accardi--Gustafson inequality (3.1), (3.2) and the triangle 
inequality (3.3)
are correct geometrically, this procedure if applied to any related inequality 
can reveal flaws in an assumed physical or
underlying probability model.  Stretching a bit, we can say this was even the
motivation of Bell in his original inequality (2.1).  From it he concluded that
hidden variables (as he perceived them) could not exist.  Also we may say that
the motivation of Accardi--Fedullo [26] was similar.  Through their inequalities
they were able to distinguish whether Kolmogorovean or quantum probability
models remained consistent with the inequalities.

What I have in mind in the present section goes somewhat beyond those 
considerations.  Here I want
to convert inequalities into equalities.  This will reveal not only violation
boundaries but also violation regions and moreover the nature of the extra terms
may conceivably be interpretted probabilistically or physically.

Consider first the Wigner inequality (2.2).  Wigner was only considering the
coplanar case in which the vectors $x,y,z$ lie in a common plane.  The Gram
determinant $G$ (3.8) vanishes if and only if the three directions are
coplanar, no matter what their frame of reference.  Then we may write the
equality in (3.8) as follows
$$(1-a_1^2 ) + (1-a_2^2 ) - (1-a_3^2 ) = 2a_3 (a_3 - a_1 a_2 )
\eqno(4.1)$$
or in the terminology of (2.2)
$$\begin{array}{ll}
&\sin^2 \left( \frac{1}{2} \theta_{12}\right) + \sin^2 \left( \frac{1}{2}
\theta_{23}\right) - \sin^2 \left( \frac{1}{2} \theta_{13} \right) \\
[1em]
&= 2\cos \left( \frac{1}{2} \theta_{13}\right) \left[ \cos \left(
\frac{1}{2}  \theta_{13} \right) - \cos \left( \frac{1}{2}
\theta_{12}\right) \cos \left( \frac{1}{2} \theta_{23} \right) \right].
\end{array} \eqno(4.2)$$
Violation of the conventionally assumed quantum probability rule $|\langle
u,v\rangle |^2 \equiv \cos^2 \theta_{u,v}$ for unit vectors $u$ and $v$
representing prepared state $u$ to be measured as state $v$ is equivalent to the
right side of (4.2) being negative.  All terms therein in principle carry
probabilistic or other physically relevant inferences.  

Williams and Clearwater \cite{willi} present essentially Wigner's formulation.
Let a Polarizer 1 with orthogonal axes
$h_1$ and $v_1$ (think:  horizontal and vertical) be inclined at angle
$\theta_1$ to the horizontal, likewise Polarizer 2 with axes $h_2$ and
$v_2$ inclined at angle $\theta_2$ to the horizontal, let $\theta_{12}$
be the angle between the polarizers.  Let $P_{xy}$ denote the probability
of detecting a photon along the $x$ and $y$ axes, respectively, of the
two detectors.  Then for a given wavefunction $\psi$ 
the measurement probabilities for the 4 possible outcomes are
$P_{v_1 v_1} = \frac{1}{2} \cos^2 \theta_{12} , P_{v_1 h_2} = \frac{1}{2}
\sin^2 \theta_{12}$,
$P_{h_1 v_2} = \frac{1}{2} \sin^2 \theta_{12} , P_{h_1 h_2} = \frac{1}{2}
\cos^2 \theta_{12}$, etc. 
Now add the third polarizer with axes $h_3$ and $v_3$ at angle $\theta_3$
to the horizontal.  Now to quote [23] ``We can write down the
following relationships from straightforward probability arguments,
$$\begin{array}{ll}
P_{v_1 h_2} &= P_{v_1 h_2 v_3} + P_{v_1 h_2 h_3} \\
[1em]
P_{v_2 h_3} &= P_{v_1 v_2 h_3} + P_{h_1 v_2 h_3} \\
[1em]
P_{v_1 h_3} &= P_{v_1 v_2 h_3} + P_{v_1 h_2 h_3}\end{array} \eqno(4.3)$$
$\cdots$.  From these relations it follows that
$P_{v_1 h_2} \geqq P_{v_1 h_2 h_3}$ and
$P_{v_2 h_3} \geqq P_{v_1 v_2 h_3}$ 
from which it follows
$P_{v_1 h_2} + P_{v_2 h_3} \geqq P_{v_1 h_2 h_3} + P_{v_1 v_2 h_3}$
or more simply
$$P_{v_1 h_2} + P_{v_2 h_3} \geqq P_{v_1 h_3} \eqno(4.4)$$
which is Bell's Inequality''.  (4.4) is in fact exactly (2.2).  Elsewhere 
\cite{gust02} I
have commented how (4.3) assumes a probability model (axiom of composite
probabilities)in which disjoint events have additive probabilities.  The last
step to (4.4) also assumes this, constituting the error.  Accardi's 
\cite{accardi1}    
fundamental paper already identified conventional probabilistic model errors
such as these.  The inequality equality (4.2) goes further and provides all the
details of the error.

As a second example, let me recall the CHSH inequality derived in (2.3)--(2.5).
Wishing now to preserve equality therein so that we may analytically
express what we may call the `violation boundaries, violation regions',
starting from (2.5) we have
$$\begin{array}{ll}
|{\mathbf a}\cdot{\mathbf b} + {\mathbf a}\cdot{\mathbf c} + {\mathbf d}
\cdot{\mathbf b} - {\mathbf d}\cdot {\mathbf c}| &= |{\mathbf a}\cdot
({\mathbf b}+{\mathbf c}) + {\mathbf d}\cdot ({\mathbf b}-{\mathbf
c})|\\
[1em]
&= |\Vert b+c\Vert \cos \theta_{a,b+c} + \Vert b-c\Vert \cos
\theta_{d,b-c}|\\
[1em]
&= |(2+2\cos \theta_{bc})^{1/2} \cos \theta_{a,b+c}\\
&\quad + (2-2\cos
\theta_{bc})^{1/2} \cos \theta_{d,b-c}| \end{array} \eqno(4.5)$$
Squaring this expression and writing everything quantum
trigonometrically,
$$\begin{array}{ll}
|{\mathbf a}\cdot{\mathbf b} + {\mathbf a}\cdot{\mathbf c} + {\mathbf
d}\cdot {\mathbf b} - {\mathbf d}\cdot {\mathbf c}|^2 &= (2+2\cos
\theta_{bc})\cos^2 \theta_{a,b+c} \\
&\quad + (2-2\cos\theta_{bc})\cos^2 \theta_{d,b-c)}\\
&\quad +2(4-4\cos^2 \theta_{bc})^{1/2} \cos \theta_{a,b+c} \cos
\theta_{d,b-c}\\
&= 4\cos^2 (\theta_{bc} /2)\cos^2 \theta_{a,b+c} \\
&\quad + 4\sin^2 (\theta_{bc}/2)\cos^2 \theta_{d,b-c}\\
&\quad + 4\sin^2 \theta_{bc} \cos \theta_{a,b+c} \cos \theta_{d,b-c}.
\end{array} \eqno(4.6)$$
In the above I used two standard trigonometric halfangle formulas.  Now
substituting the standard double angle formula $\sin \theta_{bc} = 2 \sin
(\theta_{bc} /2)\cos (\theta_{bc} /2)$ into the above we arrive at
$$|{\mathbf a}\cdot{\mathbf b} + {\mathbf a}\cdot{\mathbf c} + {\mathbf
d}\cdot{\mathbf b} - {\mathbf d}\cdot{\mathbf c}|^2 = 4[\cos \theta_{bc}
/2)\cos \theta_{a,b+c} + \sin (\theta_{bc} /2)\cos \theta_{d,b-c}]^2
\eqno(4.7)$$
and hence the {\it quantum inequality equality}
$$|{\mathbf a}\cdot{\mathbf b} + {\mathbf a}\cdot{\mathbf c} + {\mathbf
d}\cdot{\mathbf b} - {\mathbf d}\cdot{\mathbf c}| = 2|\cos (\theta_{bc}
/2)\cos \theta_{a,b+c} + \sin (\theta_{bc} /2)\cos \theta_{d,b-c}|
.\eqno(4.8)$$
We may also write the righthand side of (4.8) as twice the absolute value
of the two-vector inner product
$${\mathbf u}_1\cdot{\mathbf u}_2 \equiv (\cos (\theta_{bc} /2), \sin
(\theta_{bc} /2))\cdot(\cos \theta_{a,b+c} , \cos \theta_{d,b-c})
\eqno(4.9)$$
to arrive at the inequality equality
$$|{\mathbf a}\cdot{\mathbf b} + {\mathbf a}\cdot{\mathbf c} + {\mathbf
d}\cdot{\mathbf b} - {\mathbf d}\cdot{\mathbf c}| = 2 (\cos^2
\theta_{a,b+c} + \cos^2 \theta_{d,b-c})^{1/2} |\cos \theta_{u_1 ,u_2}|
.\eqno(4.10)$$
The right sides of these two equalities (4.8), (4.10) isolate the
``classical probability factor'' 2 from the second factor, which may
achieve its maximum $\sqrt 2$.  That the latter maximum is consistent
with the third factor in (4.10) also achieving its maximum value 1 may be
seen as follows.  Fix any directions {\bf b} and {\bf c}.  Then choose
{\bf a} relative to ${\mathbf b} + {\mathbf c}$ and choose $d$ relative
to ${\mathbf b}-{\mathbf c}$ so that $\cos^2 \theta_{a,b+c} = 1$ and
$\cos^2 \theta_{d,b-c} = 1$, respectively.  Now we may choose the free
directions {\bf b} and {\bf c} to maximize the third factor to $\cos
\theta_{u_1 ,u_2} = \pm 1$.  But that means the two-vectors ${\mathbf
u}_1$ and ${\mathbf u}_2$ are colinear and hence
$$\begin{array}{ll}
{\mathbf u}_1 &= (\cos (\theta_{bc} /2), \sin (\theta_{bc} /2)) =
2^{-1/2} (\cos \theta_{a,b+c} ,\cos \theta_{d,b-c})\\
[1em]
&= 2^{-1/2} (\pm 1, \pm 1)\end{array} \eqno(4.11)$$
and thus the important angle $\theta_{bc}$ is seen to be $\pm \pi /2$.
More to the point, the above inequality equality allows one to exactly trace
out the ``violation regions'' analytically in terms of the trigonometric
inner product condition $1 \leqq |{\mathbf u}_1\cdot {\mathbf u}_2|
\leqq \sqrt 2$.  From this point of view, there are no Bell inequalities.
Each should be replaced with an inequality equality.

Let me summarize the above.  One started with a classical probability
correlation definition (2.3) and derived a Bell inequality $|\cdots |
\leqq 2$.  The ``equality'' version of this classical probability version
would be in the individual terms
$$|v_i (a) (w_i (b)+w_i (c))+v_i (d)(w_i (b)-w_i (c))| = 2 \eqno(4.12)$$
On the other hand, inserting the quantum correlation definition into the
left side of (2.4) results (2.5) in the Bell inequality $|\cdots| \leqq
2\sqrt 2$.  My equality version (4.10) of this becomes the vector
trigonometric identity
$$|\cos \theta_{ab} +\cos \theta_{ac} +\cos \theta_{bd} -\cos
\theta_{dc}| = 2(\cos^2 \theta_{a,b+c} +\cos^2 \theta_{d,b-c})^{1/2}
|\cos \theta_{u_1 ,u_2}|.\eqno(4.13)$$
It could be useful to call (4.13) the quantum spin correlation
identity.  But it is really a new mathematical result in vector trigonometry, 
independent of any physical assumptions.

\section{Concluding Remarks}
In spite of the huge literature on physical and metaphysical interpretations of
Bell's inequalities and related theory and experiments, the operator
trigonometry provides a new and correct mathematical setting for much of that
theory.  I would assert that one correct
physical understanding of the Bell inequalities is that of basic Hilbert
space geometry, more specifically, the geometry of Euclidean and Unitary
spaces, more specifically, that of a classical but new vector
trigonometry.

The second comment is that the principal connection to physics in the 
above development is our belief
that quantum correlations are given by the quantum probability rule:  for
two normalized vectors {\bf u} and {\bf v},  the probability that a
quantum system prepared in state {\bf u} will successfully pass a test
for state {\bf v} is $|{\mathbf u}\cdot {\mathbf v}|^2 \equiv \cos^2 
\theta_{u,v}$.  The
quantum probability rule generally states that the
expectation value of an observable $A$ which has been determined
experimentally as the arithmetic mean $\langle A\rangle$ of a large
number of trials, should correspond theoretically to $Tr (AW)$ where $W$
is the statistical operator describing the state of the system.  For pure
states this quantum probability rule becomes, operationally and loosely:
the expected value is the projection onto the state.  For the spin zero
singlet state in the Bell situation the expected correlation
value is $E(a,b) = -{\mathbf a}\cdot{\mathbf b} = -\cos
\theta_{ab}$.  From this ansatz alone and my inequality equalities above, one
divides vectors {\bf a}, {\bf b}, {\bf c}, {\bf d} into `satisfaction'
and `violation' regions in whatever Hilbert space you want to take your
direction vectors from.

From this viewpoint, I would prefer that the multitude of physical
experiments over the years since [8] which have found various
physical quantum mechanical configurations in which ``Bell's 
inequality'' is violated, be restated as showing that my inequality equalities
such as (4.13)
are achieved by those physical configurations for which the right hand
side is between2 and $2\sqrt 2$.  But we know the latter is just vector
geometry.  So what these physical experiments really have shown is
various verifications of the quantum probability rule.  To repeat and
indeed overstate my point, rather than seeking ``Bell inequality
violations'', it would be more interesting to seek ``quantum probability
rule violations''.  This, because the quantum probability rule is a
far-reaching assumption, an ansatz, which in the sense of my presentation
in this paper, reduces much of quantum mechanics to a new vector
trigonometry.  Thus for example one could seek some quantum physical situation 
which could result in physical measurements for which there obtains a
right-hand-side greater than $2\sqrt 2$.  Also one could run a large number of
tests to determine what statistical distribution of occurrences manifests itself
in the region between 2 and $2\sqrt 2$.
 
As a third comment, let me make an 
assertion which seems relevant in view of the
developments of this paper:  one cannot argue either locality or
nonlocality on the basis of satisfaction or violation of Bell's
Inequality.  Bell's Inequality, notwithstanding the key and very
important role it has played in the evolving scientific revolution of
quantum mechanics, is seen in retrospect as a ``red herring'':  a
diversion distracting attention away from the real issue.  Unlike
political red herrings, the original intent of Bell and consequent investigators
was genuine.  However, from my viewpoint, the real issue as concerns nonlocality
in quantum physics is the projection rule.  This ``probability'' rule is
fundamental to Von Neumann quantum mechanics.  It is also fundamental to my
quantum trigonometry.  It is surely true for the latter, i.e., geometrically.  Is
it true for the former?

How do various authors argue nonlocality on the basis of Bell's inequalities?  B.
d'Espagnat [19, p. 124] makes it clear:  ``the violation of either the
compound premise `local causality and free will'---that is, the premises of Bell
2---or the premises of Bell 1 or Bell 3.''  Thus in particular, violation of the
CHSH inequality (2.4).  But I have shown that an occurrence yielding a (4.13)
value between 2 and $2\sqrt 2$ just means you are in that (acceptable) region of
Hilbert space.  There is no way to equate that geometric region with any concept
of nonlocality.

Maudlin \cite{maudlin} discusses nonlocality from several points of view.
However his preferred view is that of entanglement:  after the simultaneous
emission of a photon pair ``the photons are perfectly correlated:  each does what
the other does.''  There is no attenuation over distance of this quantum effect.
``Finally, the speed of quantum communication appears to be incompatible with
relativistic space--time structure.''  Most useful is his Notes [18,
p. 49] with citations about varous senses of `locality'.  Finally we come to
[18, p. 154] ``Reliable violations of Bell's inequality need not
allow superluminal signalling but they do require superluminal causation.''
While such may or may not turn out to be the case, and while my view depends in
its details upon exactly which Bell's inequality you are discussing, `reliable'
violation of Bell's inequality in my theory is not a violation.  It is just an
occurrence fitting into the larger geometric inequality--equality.

The extensive analysis [15] of the EPR paradox also considers many aspects of
`realism', violation of Bell's inequalities in a number of experimental
situations, nonlocality.  Generally when one reads the literature one finds it
easier to define locality than nonlocality.  Generally locality is thought of
[15, p. 35] as ``the intensity of interacton between objects depends inversely on
their separation'' and quantum nonlocality as [15, p. 195] long-range interference
effects.  It seemed to me that Afriat and Selleri [15] were careful not to equate
nonlocality just to violation of a Bell's inequality.  On the other hand they
state [15, p. 237] ``As is well known, any proof of Bell's Theorem has two steps.
The first step is a deduction of inequalities from the assumptions of reality and
locality, and the second the exhibition of a contradiction with quantum
mechanics.''   My point of view is that such inequalities are a direct
mathematical consequence of Hilbert space structure, independent of any physical
ideas.  

Peres [12, pp. 169, 173] fully accepts the traditional Bell view.  ``In summary,
there is no escape from nonlocality.  The experimental violation of Bell's
inequality leaves only two logical possibilities:  either some simple physical
systems$\cdots$are essentially nonlocal, or it is forbidden to consider
simultaneously the possible outcomes of mutually exclusive experiments.''  Again,
not having my operator trigonometric view of the geometry of quantum
probabilities does not permit a full understanding of Bell's inequalities.
\newpage

Van Frassen \cite{frassen} emphasizes the philosophical and logic aspects of
quantum interpretation.  Generally cautious in stating what may be deduced,
nonetheless he states Bell's inequalities [22, pp. 93, 104, 348] in
the form $p_{12} + p_{23} \geqq p_{13}$ of Wigner [9].  As I have shown here,
these inequalities (2.2) and (4.4) are flawed both quantum probabilistically and
geometrically.  The trouble with empirical positivism (i.e., the philosophical
school of Carnap and others) is that it (in my opinion) permits to obviate
searching for or investigating mathematically deeper fundamental physical
meanings.  On the other hand that philosophy does lead one back to Von Mises view
of probabilities, that they best be given by relative frequencies.

Finally, there is another development which serves to  bring into question the
Von Neumann projection rule.  That is the Zeno's paradox, which I treat in 
\cite{gustto}.
I cannot go into all the citations and the mathematical details here but I show
in \cite{gustto} that there are serious unresolved mathematical issues about the
`continuous observation' operator limit $s-\lim_{n\rightarrow\infty} (PU(t/n)
P)^n$.  As a consequence in [50] I propose what I call domain-compatible
measuring operators to replace the projections $P$.

Thus in summary one may say that Bell, Wigner, others, focused on physical
axioms which led them to certain inequalities.  Accardi, Fedullo, others focused
on probabilistic axioms which led them to certain inequalities.  I have focused
on geometrical axioms which have led me to certain inequalities.  Moreover I have
extended the notion of such ``Bell'' inequalities to what I called here
``inequality--equalities'' which may then be used to reflect back more accurate
meanings of the earlier inequalities.

\section{Acknowedgements}
I was led into these matters by Luigi Accardi following a discussion with him at
Les Treilles, France in July 1996.  After that discussion, on my own I found the
key paper [26] and saw the coincidence (3.1), (3.2).  I have since enjoyed
several profitable discussions with Accardi at meetings.  Also I would like to
mention useful conversations on this subject with Andrei Khrennikov, Harald
Atmanspacher, Han Primas, and Urs Wild.

\end{document}